\documentclass{cjpsuppl}
\def\mevc {\ifmmode {\rm MeV}/c \else MeV$/c$\fi}
\def\mevcc {\ifmmode {\rm MeV}/c^2 \else MeV$/c^2$\fi}
\def\gevc {\ifmmode {\rm GeV}/c \else GeV$/c$\fi}
\def\gevcc {\ifmmode {\rm GeV}/c^2 \else GeV$/c^2$\fi}
\def\ra   {\rightarrow}
\newcommand{\Bs} {\ifmmode B_{\mbox{\sl s}}^{0}
                       \else $B_{\mbox{\sl s}}^{0}$\fi}
\newcommand{\Ds} {\ifmmode D_{\mbox{\sl s}}^{+}
                       \else $D_{\mbox{\sl s}}^{+}$\fi}
\newcommand{\dms} {\ifmmode \Delta m_{\mbox{\sl s}} \else 
                           $\Delta m_{\mbox{\sl s}}$\fi}
\newcommand{\xs} {\ifmmode x_{\mbox{\sl s}} \else 
                           $x_{\mbox{\sl s}}$\fi}
\input{epsf}
\begin{document}
\title{\boldmath{$B$}~Physics at the Tevatron}
\authori{Manfred Paulini \\
(Representing the CDF and D\O\ Collaboration)}
\addressi{Department of Physics, Carnegie Mellon University,
Pittsburgh, PA 15213, USA}
\authorii{}    \addressii{}
\authoriii{}   \addressiii{}
\authoriv{}    \addressiv{}
\authorv{}     \addressv{}
\authorvi{}    \addressvi{}
\headtitle{$B$~Physics at the Tevatron}
\headauthor{M.~Paulini}
\lastevenhead{M.~Paulini: $B$~Physics at the Tevatron}
\refnum{}
\daterec{20 Sept 2003}
\suppl{A}  \year{2003} \setcounter{page}{1}
\maketitle

\begin{abstract}
After a five year upgrade period, the Fermilab experiments
CDF and D\O\ are taking high quality data in Run\,II
of the Tevatron Collider.
We report on the start-up of both detectors and present a selection of
first $B$~physics results from the Tevatron.
We also compare different $B$~hadron producers such as
the $\Upsilon(4S)$ with the hadron collider environment and
discuss general features of $B$~physics at a hadron collider.
\end{abstract}

\section{Introduction}

Traditionally, $B$ physics has been the domain of $e^+ e^-$ machines
operating on the $\Upsilon(4S)$ resonance or the $Z^0$ pole.
But the UA\,1~Collaboration has already shown that $B$~physics
is feasible at a hadron collider environment (see for example
Ref.~\cite{bfeasi}). 
The first signal
of fully reconstructed $B$~mesons at a hadron 
collider has been published by the CDF~Collaboration in 
1992~\cite{cdf_firstB}. 
CDF reconstructed a handful of $B^+ \ra J/\psi K^+$
events in a data sample of 2.6~pb$^{-1}$ taken during the Tevatron
Run\,0 at the end of the 1980's. Since then experimental techniques improved
significantly, 
especially with the development of high precision silicon vertex
detectors. 

The CDF and D\O~experiments can look back to an already
successful $B$~physics program during 
the 1992-1996 Run\,I data taking period (for a review of $B$~physics
results from, for example, CDF in Run\,I see Ref.~\cite{myrevart}).  
Nowadays, $B$~physics results from a hadron collider are fully
competitive with the $e^+ e^-$ $B$~factories.
As discussed later in this review, 
with the operation of a hadronic track trigger, CDF 
reconstructs fully hadronic $B$~decay modes without leptons in the
final state. In many cases, the measurements performed at
the Tevatron Collider are complementary to the $B$~factories. For
example, no $\Bs$~mesons or baryons containing $b$~quarks are produced
on the $\Upsilon(4S)$~resonance.

\section{Status of Tevatron Run\,II}

\subsection{The Upgraded Tevatron Accelerator}

The Fermilab accelerator complex has undergone a major upgrade in
preparation for Tevatron Run\,II.
The centre-of-mass energy has been increased to 
1.96~TeV as compared to 1.8~TeV during Run\,I. 
But most importantly, the Main Injector, a new 150~GeV
proton storage ring, has replaced the Main Ring as injector of protons and
anti-protons into the Tevatron. The Main Injector also 
provides higher proton intensity onto the anti-proton production target,
allowing for a luminosity goal of 
1-2$\times 10^{32}$~cm$^{-2}$s$^{-1}$, representing a luminosity  increase
of more than an order of magnitude.
The present bunch crossing time
is 396~ns with a $36\times36$ $p\bar p$ bunch operation.
An upgrade to a 132~ns bunch crossing time has been indefinitely
postponed. 
The luminous region of the Tevatron beam 
has an RMS of $\sim\!30$~cm~along the beamline ($z$-direction)
 with a transverse beamwidth of about 25-30~$\mu$m. 

The initial Tevatron luminosity steadily increased from 2002 to 2003. 
By the summer of 2003, the peak luminosity reached 
by the Tevatron is $\sim\!5\cdot 10^{31}$~cm$^{-2}$s$^{-1}$.
The total integrated luminosity delivered by the Tevatron to CDF and
D\O\ by the time of this conference 
is $\sim\!270$~pb$^{-1}$.
About 200~pb$^{-1}$ were recorded to tape by each CDF and D\O.
For comparison, the Run\,I data set comprised
110~pb$^{-1}$.
However, most results
shown in this review use about 70-110~pb$^{-1}$ of data. The data
taking efficiency of both collider experiments has reached about
80-95\% on average. 

\subsection{CDF Detector Performance in Run\,II}

The CDF detector improvements for Run\,II~\cite{cdfup} were motivated by
the shorter 
accelerator bunch spacing of up to 132~ns and the increase in luminosity by
an order of magnitude. All front-end and trigger electronics has been
significantly redesigned and replaced. A DAQ upgrade allows the operation
of a pipelined trigger system. CDF's tracking 
system was completely renewed. It consists of  
a new Central Outer Tracker (COT) with 30\,200 sense wires
arranged in 96 layers combined into four axial and four stereo
superlayers. It also provides d$E$/d$x$ information for particle
identification. 
The Run\,II silicon vertex
detector consists of seven double sided layers and one single sided layer
mounted on the beampipe covering a total radial area from 1.5-28~cm. The
silicon vertex detector covers the full Tevatron luminous 
region and allows
for standalone silicon tracking up to a pseudo-rapidity $|\eta|$ of 2. The
forward calorimeters have been replaced by a new scintillator tile based
plug calorimeter which gives good electron identification up to 
$|\eta|=2$.
The upgrades to the muon system almost double the central
muon coverage and extend it up to $|\eta|\sim1.5$.

The most important improvements for $B$~physics in Run\,II are a 
Silicon Vertex Trigger (SVT)
and a Time-of-Flight (ToF) system with a resolution of about
100~ps. The later employs 216 three-meter-long
scintillator bars located between the outer radius of the COT
and the superconducting solenoid.
The Time-of-Flight system will be most beneficiary for the identification
of kaons with a 2\,$\sigma$-separation between $\pi$ and $K$ for
$p<1.6$~\gevc. The SVT is discussed in more detail in
Sec.~\ref{sec:svt}.

\subsection{The Upgraded D\O\ Detector}

The D\O~detector also went through a major upgrade before the beginning of
Run\,II~\cite{dup}. The inner tracking system was completely 
replaced and includes a new Silicon tracker surrounded by a
Scintillating Fiber tracker, both of which are enclosed in a 2~Tesla
solenoidal magnetic field. Pre-shower counters are located before the
uranium/liquid-argon calorimeter to improve the electron and photon
identification. The already excellent muon system has been further improved
by adding more shielding to reduce beam background.

The Run\,II D\O\ detector has excellent tracking and lepton acceptance. Tracks
with pseudo-rapidity as large as 2.5-3.0 ($\theta \approx
10^{\circ}$) and transverse momentum $p_T$ as low as 180~\mevc\ can be
reconstructed.
The muon system can identify muons within $|\eta| < 2.0$. The minimum
$p_T$ of the reconstructed muons varies as a function of $\eta$. In
most of the results presented, muons were required to have $p_T > 2$~\gevc.
Low momentum electron identification is currently limited to $p_T > 2$~\gevc\
and $|\eta| < 1.1$. However, the D\O\ Collaboration is making serious
attempts to improve both the momentum and $\eta$ coverage.

A silicon based hardware trigger is being commissioned which will
allow to trigger on long-lived particles, such as the daughters of
charm and beauty hadrons. D\O\ expects to include this trigger in
the online system by the end of 2003.
D\O\ already applies impact parameter requirements in form
of a software trigger at Level\,3.

\section{Features of \boldmath{$B$} Physics at a Hadron Collider}

In this section, we highlight some of the features of $B$~physics at a
hadron collider. We make an attempt to describe how $B$~decays are
studied at the Tevatron emphasizing some tools used to find  
$B$~decay products in hadronic collisions.

We first compare different
producers of $B$~hadrons. Table~\ref{bproducers} summarizes some of
the important features of $B$~physics experiments and the accelerators
at which they operate. There are three main approaches to producing
$B$~hadrons. First, 
$e^+e^- \ra \Upsilon(4S) \ra B\bar B$ where the $B$~factories dominate 
with the BaBar detector at SLAC and the Belle experiment at
KEK. The pioneers of $B$~physics at the
$\Upsilon(4S)$~resonance were the CLEO experiment located at the
CESR storage ring at Cornell and the ARGUS experiment at the DORIS
storage ring at DESY.
Second, $e^+e^- \ra Z^0 \ra b\bar b$ at the former LEP accelerator at
CERN, as well as the SLD detector at the SLC   
Collider at SLAC. Finally, $p\bar p \ra b\bar b X$ at the Tevatron,
where the CDF and D\O\ detectors operate. 
The main motivation for studying $B$ physics at
a hadron collider is the large $b$~quark production cross section  
$\sigma_{b}\sim\!50~\mu$b within the central detector regions
(see Table~\ref{bproducers}). 

\begin{table}[tbp]
\caption{
Comparison of important features of different experiments studying
$B$~physics.}  
\small
\begin{center}
\begin{tabular}{c|ccc} 
\hline
 & & & \\
 \vspace*{-0.6cm} \\
 & $e^+e^- \ra \Upsilon(4S) \ra B^{ }\bar B$ & $e^+e^- \ra Z^0 \ra b\bar b$ 
 & $p\bar p \ra b\bar b X$ \\ 
\hline 
 & & & \\
 \vspace*{-0.6cm} \\
Accelerator & PEP\,II, KEK, CESR & LEP, SLC & Tevatron \\
Detector & BaBar, Belle & ALEPH, DELPHI & CDF, D\O \\
 & ARGUS, CLEO & L3, OPAL, SLD &  \\
$\sigma(b\bar b)$ & $\sim\!1$ nb & $\sim\!6$ nb & $\sim\!50~\mu$b  \\
$\sigma(b\bar b):\sigma(had)$  & 0.26 & 0.22 & $\sim\!0.001$  \\ 
$B^0,\ B^+$ & yes & yes & yes \\
$\Bs,\ B_c^+,\ \Lambda_b^0$ & no & yes & yes \\
Boost $<\beta\gamma>$  & 0.06 & 6 & $\sim$\,1\,-\,4 \\
$b\bar b$ production  & both $B$ at rest & $b\bar b$ back-to-back & 	
	$b\bar b$ not back-to-back \\ 
Multiple events   & no & no & yes  \\ 
Trigger  & inclusive & inclusive & leptons, hadrons \\ 
\hline
\end{tabular}
\vspace{-1mm}
\label{bproducers}
\end{center}
\end{table}

\begin{figure}[tb]
\centerline{
\epsfxsize=6.1cm
\epsffile{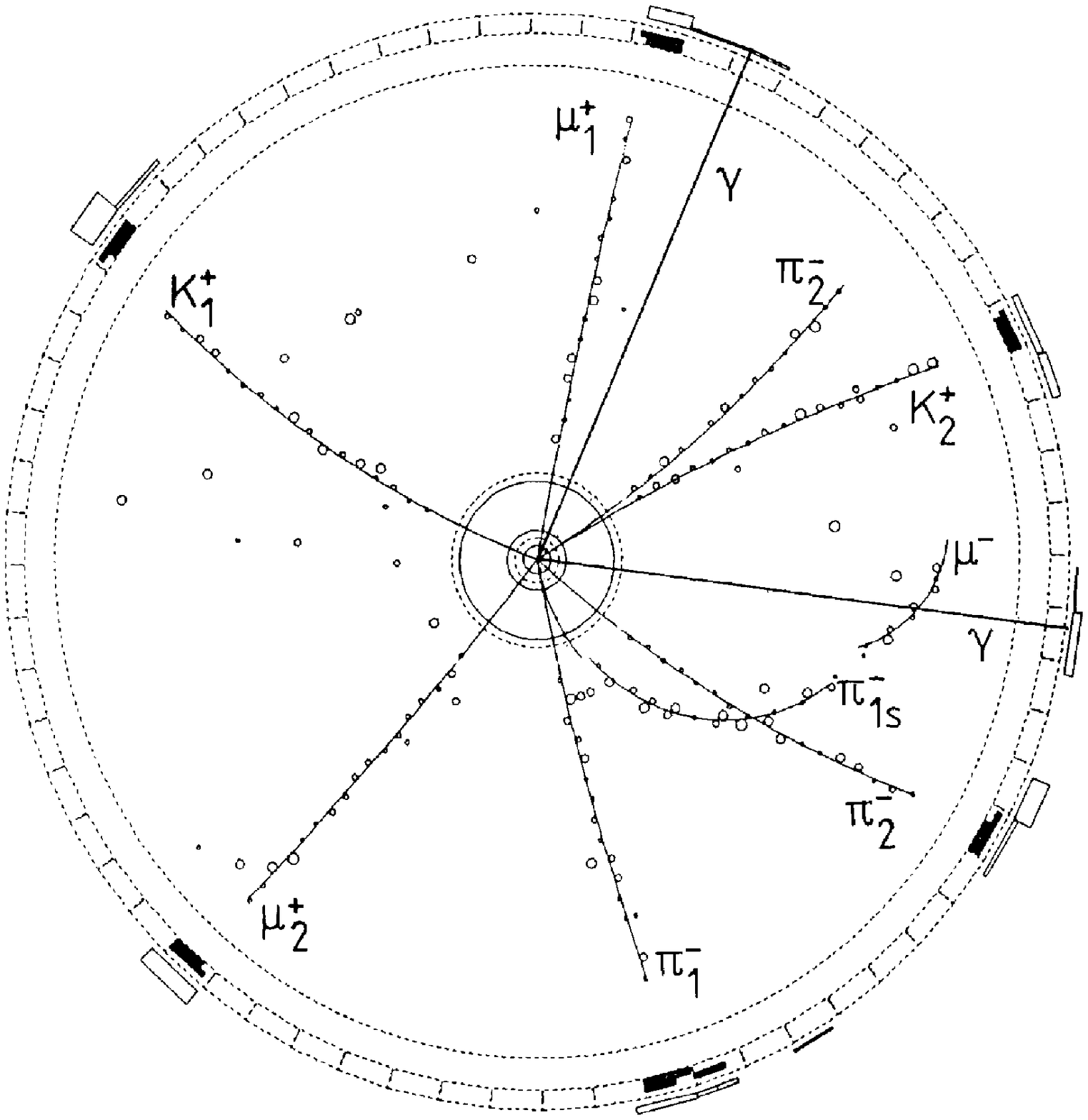}
\epsfxsize=6.1cm
\epsffile{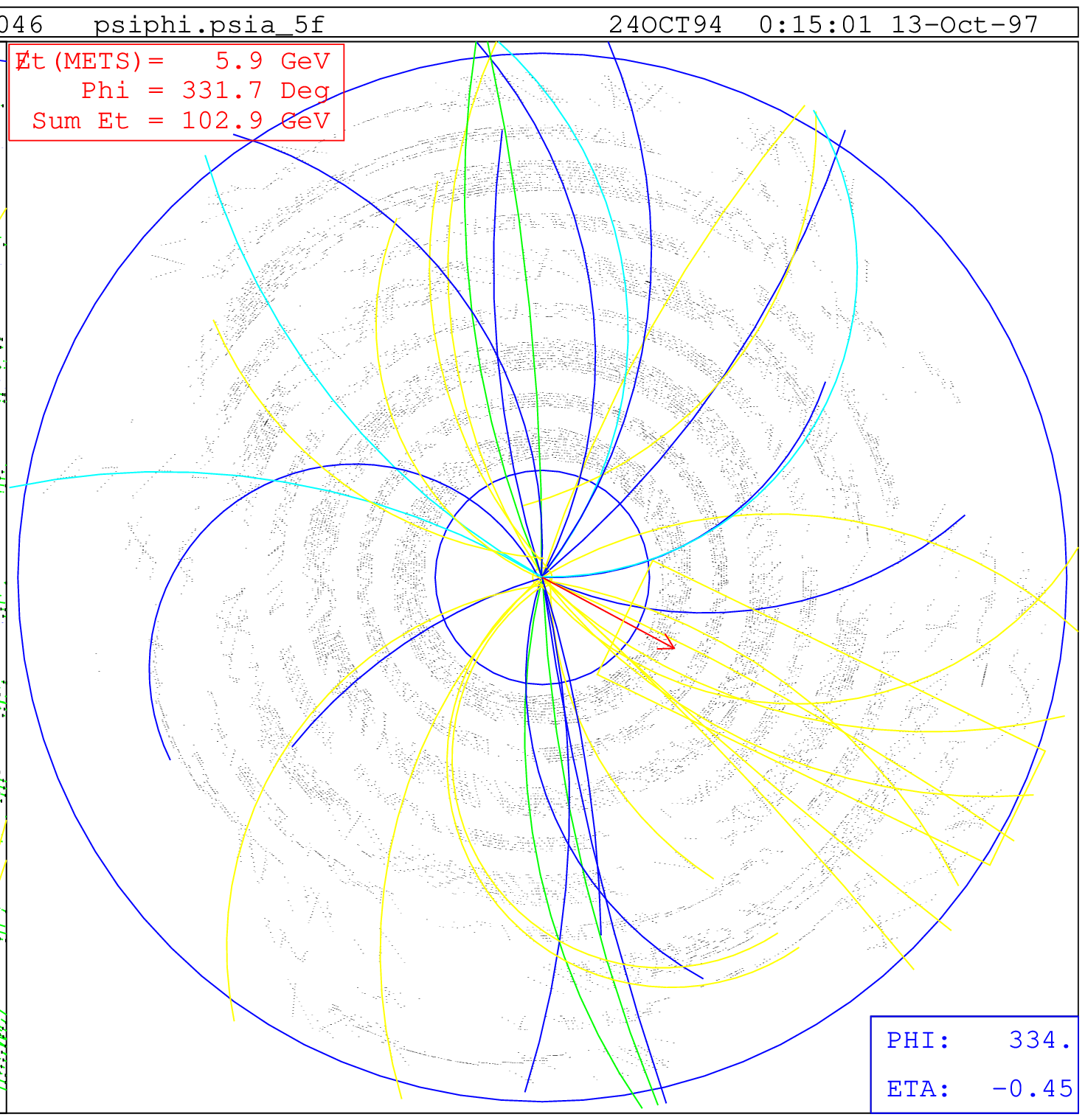}
\put(-205,160){\large\bf (a)}
\put(-28,155){\large\bf (b)}
}
\caption{
(a) A typical $B$ event at the $\Upsilon(4S)$ in the $r\,\varphi$~view from 
the ARGUS experiment. 
(b) A typical $B$ event at the Tevatron recorded with 
the CDF detector.
}
\label{evt_bs}
\end{figure}

Figure~\ref{evt_bs}(a) shows a typical $B$~event at the
$\Upsilon(4S)$ recorded with the ARGUS detector.
At the $\Upsilon(4S)$ resonance, only $B^0\bar B^0$ or
$B^+B^-$ pairs are produced nearly at rest, resulting in a spherical event
shape with an average charged particle multiplicity of about ten tracks. 
At the LEP or SLC accelerators, $b\bar b$ quark pairs are produced from
the decay of the $Z^0$~boson where both quarks share half of the energy
of the $Z^0$ resonance of 91.2~GeV. This results in two $b$~jets
being back-to-back. The
average boost of $B$~hadrons at the $Z^0$~resonance is $\beta\gamma \sim 6$. 
  
Figure~\ref{evt_bs}(b) represents a typical $B$~event from CDF.
No well-defined jet structure is visible; the average
multiplicity is about 50 charged tracks including tracks from the
``underlying event'' particles.
It might appear challenging to find the $B$~decay products in this
quite messy environment of a hadronic collision. 
One way to extract
$B$~decays in a $p\bar p$ collision is to take advantage of the
relatively long lifetime of $B$~hadrons resulting in a 
$B$~decay vertex which is clearly separated from the primary $p\bar p$
interaction vertex by hundreds of microns. 

\begin{figure}[tbp]
\centerline{
\epsfxsize=6.1cm
\epsffile{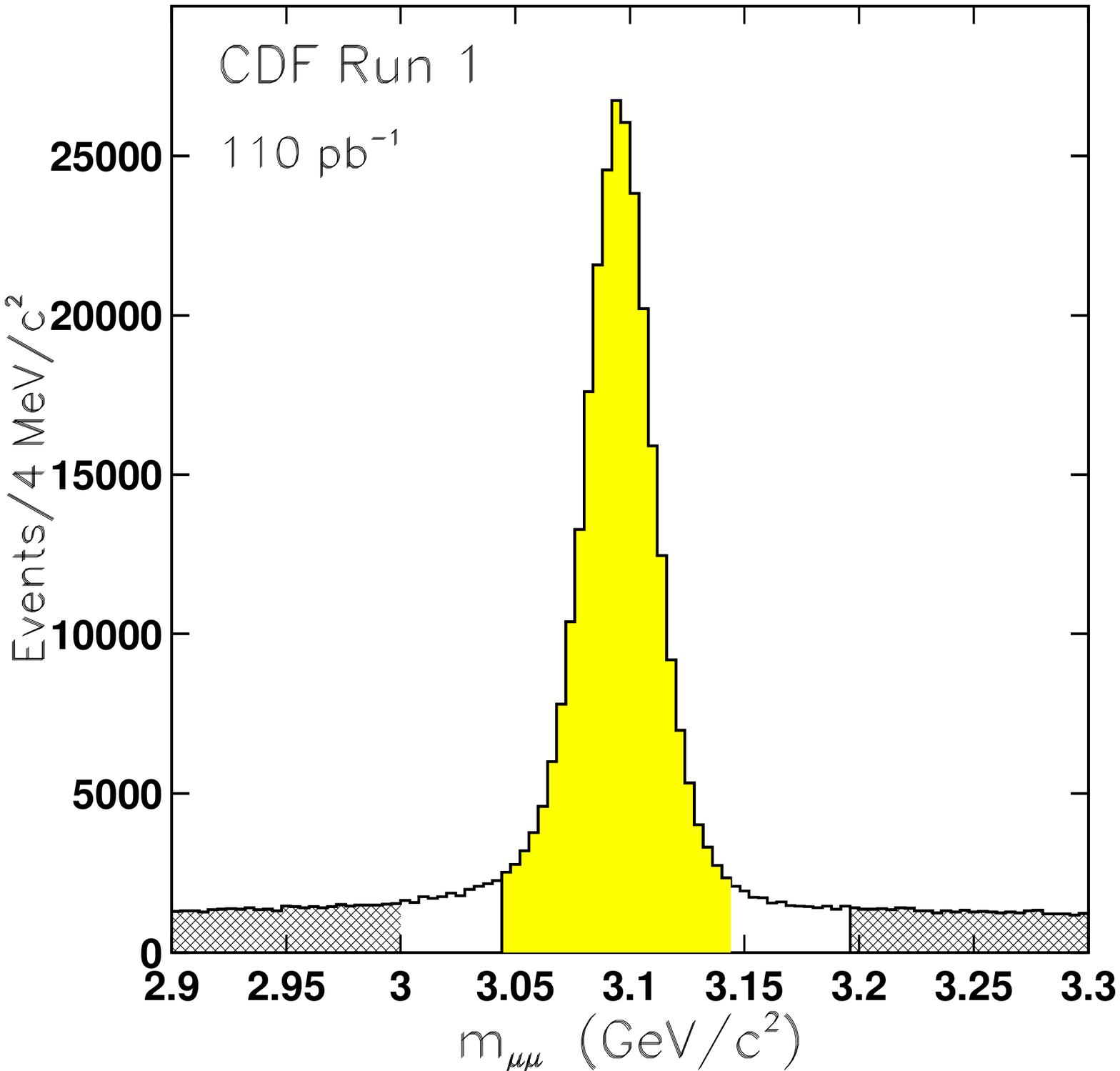}
\epsfxsize=6.1cm
\epsffile{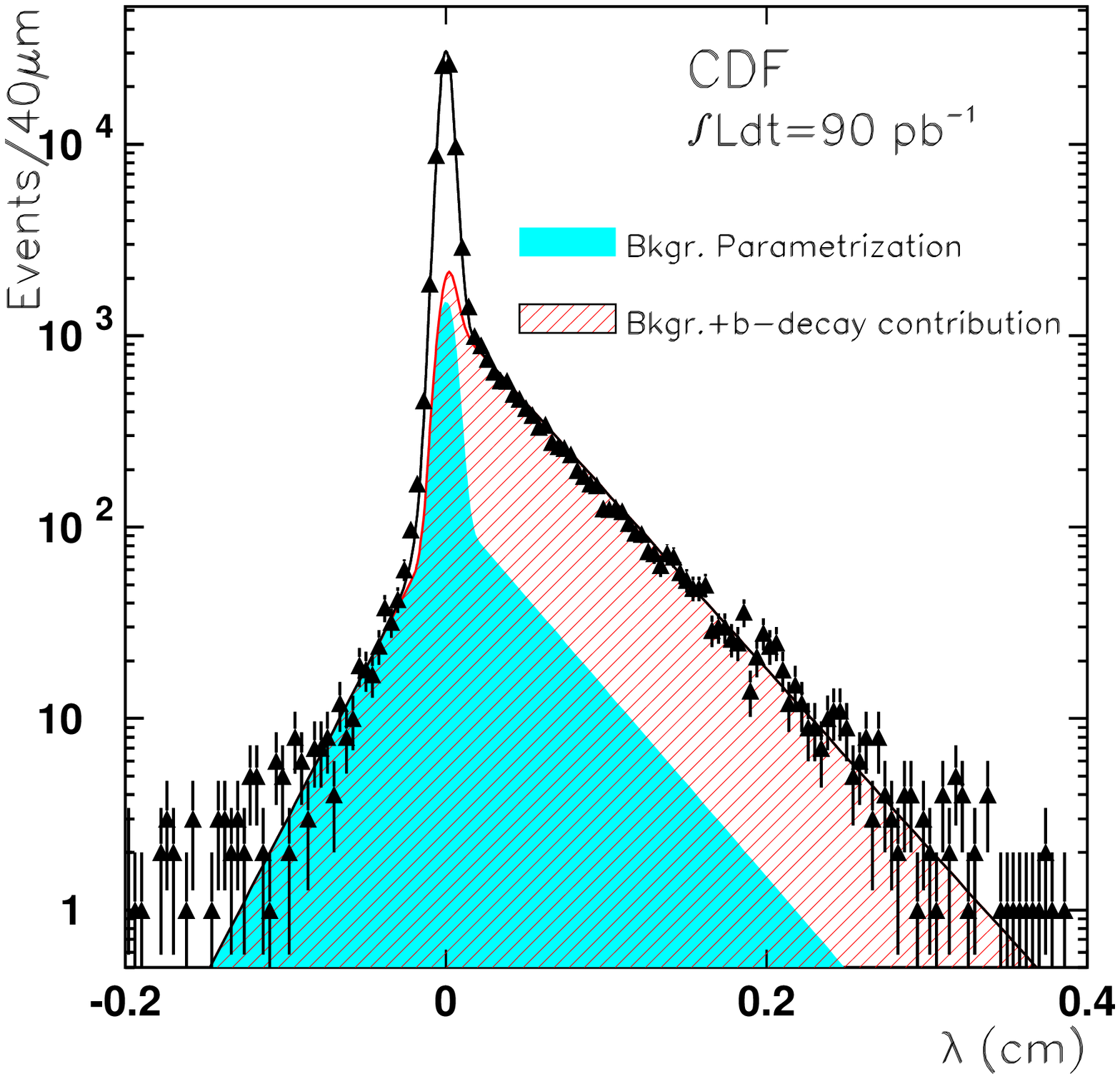}
\put(-210,143){\large\bf (a)}
\put(-145,143){\large\bf (b)}
}
\caption{
(a) Invariant mass distribution of oppositely charged muon pairs from
the CDF Run\,I dimuon trigger.
(b) Decay length distribution of the $J/\psi$~signal events.
}
\label{jpsi_mass_tau}
\end{figure}

One important feature for $B$~physics at a hadron collider is
a good tracking capability which is usually achieved with a central
tracking chamber. Together with a silicon detector assisting in
tracking,  
an excellent track momentum resolution
translates into an excellent invariant mass resolution, as illustrated
in Fig.~\ref{jpsi_mass_tau}(a). Here, the dimuon invariant mass is
displayed for muons from the Run\,I dimuon trigger stream at CDF. 
A prominent $J/\psi$ peak
is visible on low background. The mass resolution of
the $J/\psi$ peak is about 16~\mevcc. 

In addition to excellent tracking, superb vertexing is the other
essential feature of successful $B$~physics studies at a hadron collider.
This is demonstrated in Fig.~\ref{jpsi_mass_tau}(b) where the two muons of
the $J/\psi$ signal candidates [light-shaded area in
Fig.~\ref{jpsi_mass_tau}(a)] are vertexed using tracking information
from the silicon detector. The two-dimensional distance between the
primary $p\bar p$ interaction vertex and the reconstructed dimuon
vertex is plotted. This distribution shows several features: a
prominent peak at zero decay length results from prompt $J/\psi$
candidates which  
are produced at the primary interaction vertex and constitute about 80\% of all
$J/\psi$ candidates. The width of this peak reveals information about
the vertexing resolution, which is on average $40$-$50~\mu$m, for this
sample. At positive decay lengths, $J/\psi$ mesons from
$B$~hadron decays are described by an exponential slope. At a distance
of about 100~$\mu$m from the 
primary interaction vertex, mainly $J/\psi$'s from $B$~decays remain. 
There is also a small exponential slope at negative decay lengths
where the particle seems to decay before the point where
it is produced. These events result from the combinatorial background
underneath the $J/\psi$ signal. This is indicated by events from the
$J/\psi$ sidebands [dark-shaded regions in Fig.~\ref{jpsi_mass_tau}(a)]
which describe well the distribution at negative decay lengths as 
seen by the dark-shaded area in Fig.~\ref{jpsi_mass_tau}(b).

\subsection{Triggering on \boldmath{$B$}~Decay Products}
\label{sec:svt}

The total inelastic $p\bar p$ cross section at the Tevatron is 
about three orders of magnitude larger than the $b$~production cross
section. The CDF and D\O\ trigger system is therefore the most
important tool for finding $B$~decay products. In addition, the cross
section for $b$~quark 
production is steeply falling. It drops by almost two orders of magnitude
between a $b$~quark $p_T$ of about 8~\gevc\ and 25~\gevc. 
To find $B$~decay products in hadronic collisions, it is desirable
to go as low as possible in the decay products transverse
momentum, exploiting as much as possible of the steeply falling
$b$~cross section. Of course, the limiting factor is the bandwidth of
the experiment's data acquisition system. 

In Run\,I, all $B$~physics triggers at CDF and D\O\ were based on
leptons including single and dilepton triggers. 
In Run\,II, both experiments still exploit heavy flavour decays which
have leptons in the final state.  Identification of
dimuon events down to very low momentum is possible,
allowing for efficient $J/\psi \rightarrow \mu^+\mu^-$
triggers. As a consequence, both experiments are able to
fully reconstruct $B$~decay modes involving $J/\psi$'s.
Triggering on dielectrons
to isolate $J/\psi \rightarrow e^+e^-$ decays 
is also possible, although at low momentum backgrounds
become more problematic.  
CDF has implemented a 
$J/\psi\rightarrow e^+e^-$ trigger requiring each electron $p_T>2$~\gevc.  

Both experiments also use inclusive lepton triggers designed
to accept semileptonic $B\rightarrow \ell \nu_\ell X$ decays.
D\O\ has an inclusive muon trigger with excellent acceptance,
allowing them to accumulate very large samples of semileptonic
decays.  The CDF semileptonic triggers require an additional
displaced track associated with the lepton,  providing 
cleaner samples with smaller yields.

New to the CDF detector is the ability to select events
based upon track impact parameter.  
The Silicon Vertex Trigger 
gives CDF access to purely hadronic $B$~decays and makes CDF's $B$~physics
program fully competitive with the one at the
$e^+e^-$~$B$~factories. The hadronic track trigger
is the first of its kind operating successfully at a hadron collider. It
works as follows: With a
fast track trigger at Level\,1, CDF finds track pairs in the COT
with $p_T>1.5$~\gevc. At Level\,2, these tracks are
linked into the silicon vertex detector and cuts on the track impact
parameter (e.g.~$d > 100$ $\mu$m) are applied. 
The SVT track impact parameter resolution is about 50~$\mu$m including a
33~$\mu$m contribution from the transverse beam spreading.
The original motivation for CDF's hadronic track trigger was to select
$B^0 \ra \pi\pi$ decays to be used for $CP$~violation studies.

\section{Selected \boldmath{$B$} Physics Results from the Tevatron}

With the different $B$~trigger strategies above, the
Collider experiments are able to trigger and reconstruct large samples
of heavy flavour hadrons. 
To give an idea about the sample sizes available for heavy flavour
analyses,
the approximate yield 
for $D^0\ra K^-\pi^+$ is $\sim\!6000$ events per pb$^{-1}$, 
for $B^-\ra D^0\pi^-$ it is $\sim\!16$ events, 
for $J/\psi\ra \mu^+\mu^-$ it is $\sim\!7000$ events, 
for $B^-\ra J/\psi K^-$ it is $\sim\!11$ events or
for $B\ra D\ell\nu$ it is $\sim\!400$ events per pb$^{-1}$ at the Tevatron. 
In the following we discuss some selected
$B$~physics results from CDF and D\O.

\subsection{\boldmath{$B$}~Hadron Masses and Lifetimes}

Measurements of $B$~hadron masses and lifetimes are basic calibration
measures to demonstrate the understanding of heavy flavour
reconstruction. For example, CDF uses exclusive $B$~decay modes into
$J/\psi$ mesons for precision measurements of $B$~hadron masses
reconstructing the decay modes 
$B^0\ra J/\psi K^{\ast 0}$,
$B^+\ra J/\psi K^+$, $\Bs\ra J/\psi\phi$ and 
$\Lambda_b\ra J/\psi\Lambda$.
These modes combine good signal  
statistics with little background.  
The results of the mass measurements are summarized in
Table~\ref{tab:bmasslife}. 

\begin{table}[tbp]
\caption{
Summary of $B$~hadron mass and lifetime measurements from CDF and D\O.
}
\small
\begin{center}
\begin{tabular}{cccc}
\hline
Mode & Mass (CDF) [\mevcc] & Lifetime (CDF) [ps] & Lifetime (D\O) [ps] \\
\hline
$B^0\ra J/\psi K^{\ast 0}$ & $5280.30\pm0.92\pm0.96$ & 
$1.51\pm0.06\pm0.02$ & $1.51\pm0.18\pm0.20$ \\
$B^+\ra J/\psi K^+$  & $5279.32\pm0.68\pm0.94$ & 
$1.63\pm0.05\pm0.04$ & $1.65\pm0.08^{+0.09}_{-0.12}$ \\
$\Bs\ra J/\psi\phi$ & $5365.50\pm1.29\pm0.94$ & 
$1.33\pm0.14\pm0.02$ & $1.19\pm0.18\pm0.14$ \\
$\Lambda_b\ra J/\psi\Lambda$ & $5620.4\pm1.6\pm1.2$ & 
$1.25\pm0.26\pm0.10$ & $-$ \\
\hline
\end{tabular}
\vspace{-1mm}
\label{tab:bmasslife}
\end{center}
\end{table}

The proper time of a $B$ decay is determined from the distance between
the primary vertex of the $p\bar p$~collision and the $B$ meson decay
vertex measured in the plane transverse to the beam axis: 
$L^B_{xy} = (\vec{x}_B - \vec{x}_{prim})\cdot \vec{p}_T/|\vec{p}_T|$,
where $\vec{p}_T$ is the measured transverse momentum vector.
At D\O, the primary vertex is reconstructed individually for each event.
The typical resolution of $L^B_{xy}$ is 40~$\mu$m.
CDF uses the run-averaged beam position whose contribution to the
$L^B_{xy}$ uncertainty is $\sim\!30~\mu$m.

\begin{figure}[tbp]
\centerline{
\epsfxsize=6.2cm
\epsffile{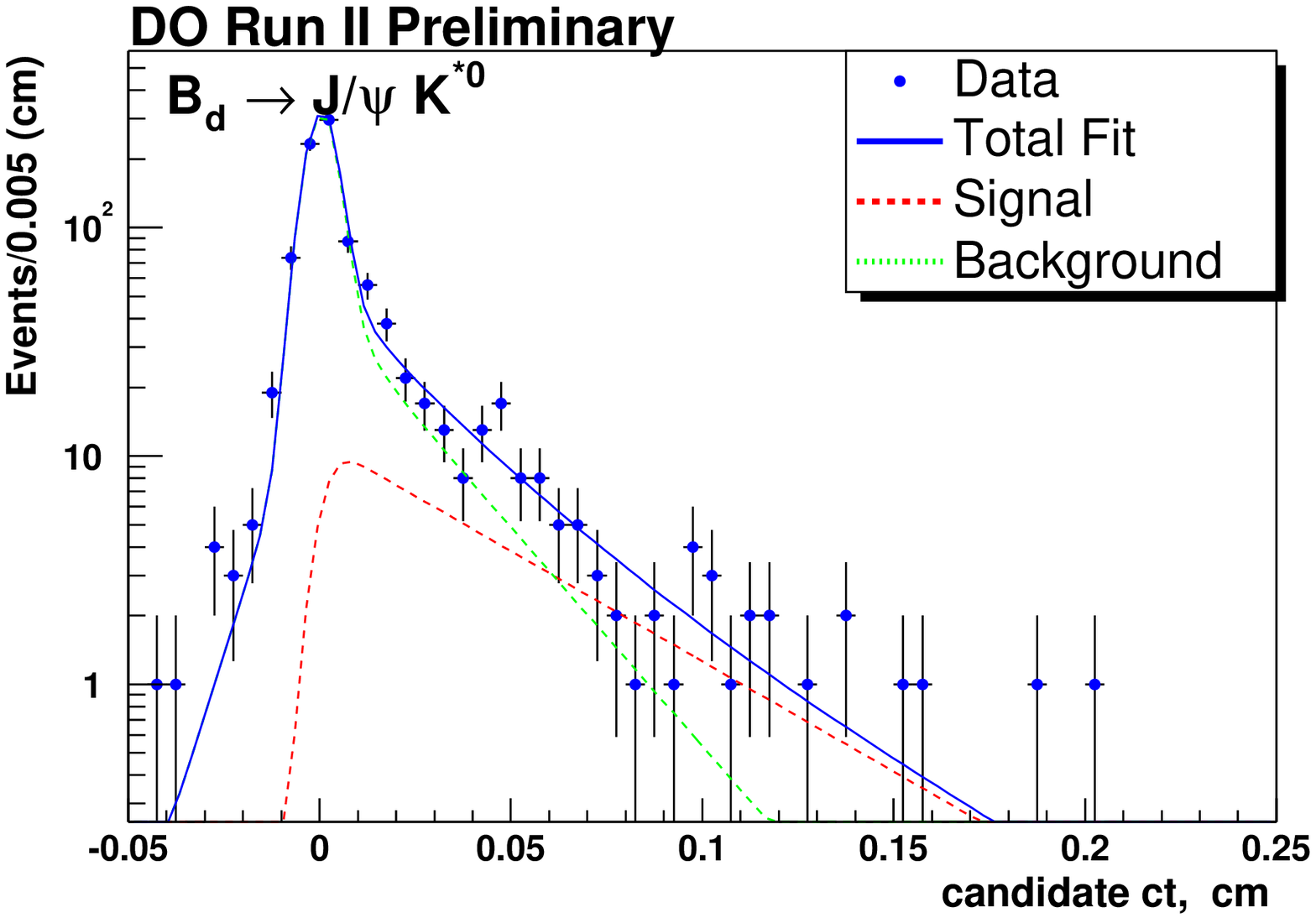}
\epsfxsize=6.1cm
\epsffile{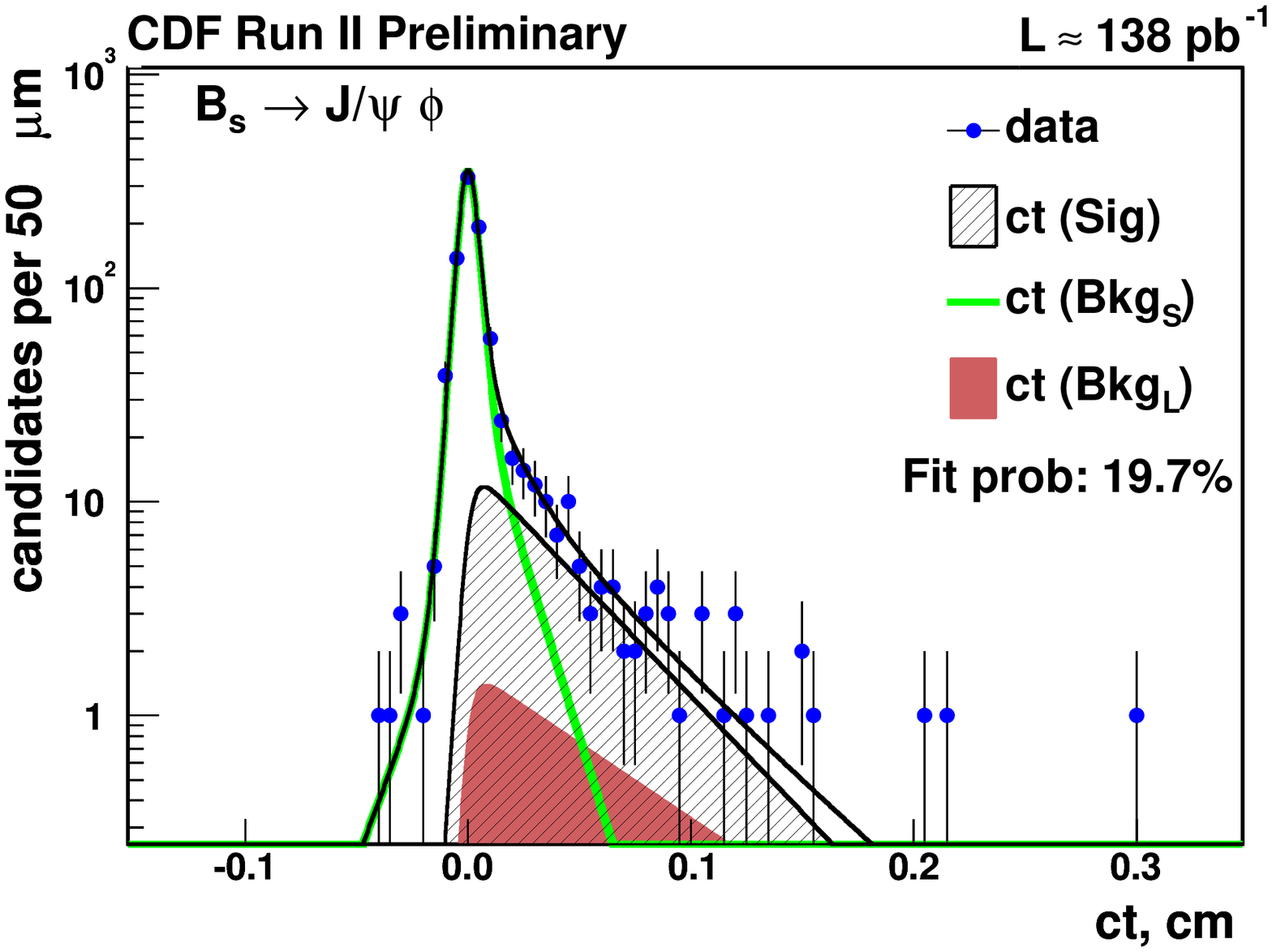}
\put(-265,95){\large\bf (a)}
\put(-145,95){\large\bf (b)}
}
\caption{
Examples of proper decay length distribution for 
(a) $B^0 \rightarrow J/\psi K^{\ast 0}$ at D\O\ and 
(b) $\Bs \rightarrow J/\psi \phi$ at CDF.
}
\label{fig:blifepsi}
\end{figure}

In the case of fully reconstructed $B$ hadron decays, the proper  lifetime
$\tau$ is obtained by $c\tau = L^B_{xy}\cdot M_{B}/p_T$,
where $M_{B}$ is the $B$ hadron mass. 
In  the case of inclusive decays where the $B$~hadron is not fully
reconstructed, a boost correction obtained  from a Monte Carlo
simulation is usually applied.
Both experiments have measured the lifetimes of the 
$B^+$, $B^0$ and $\Bs$ mesons from the decay channels
$B^+ \rightarrow J/\psi K^+$, $B^0 \rightarrow J/\psi K^{\ast 0}$
and $\Bs \rightarrow J/\psi \phi$, respectively.
Examples of proper decay length distributions and fit results are shown
in Figure~\ref{fig:blifepsi} for the decay mode $J/\psi K^{\ast 0}$ at D\O\ and 
$J/\psi \phi$ at CDF. The resulting lifetime measurements from CDF and
D\O\ are summarized in Table~\ref{tab:bmasslife} which also includes
preliminary measurements of the $\Lambda_b\ra J/\psi\Lambda$ mass and
lifetime from CDF.

\subsection{Prompt Charm Cross Section}

\begin{figure}[tbp]
\centerline{
\epsfxsize=6.1cm
\epsffile{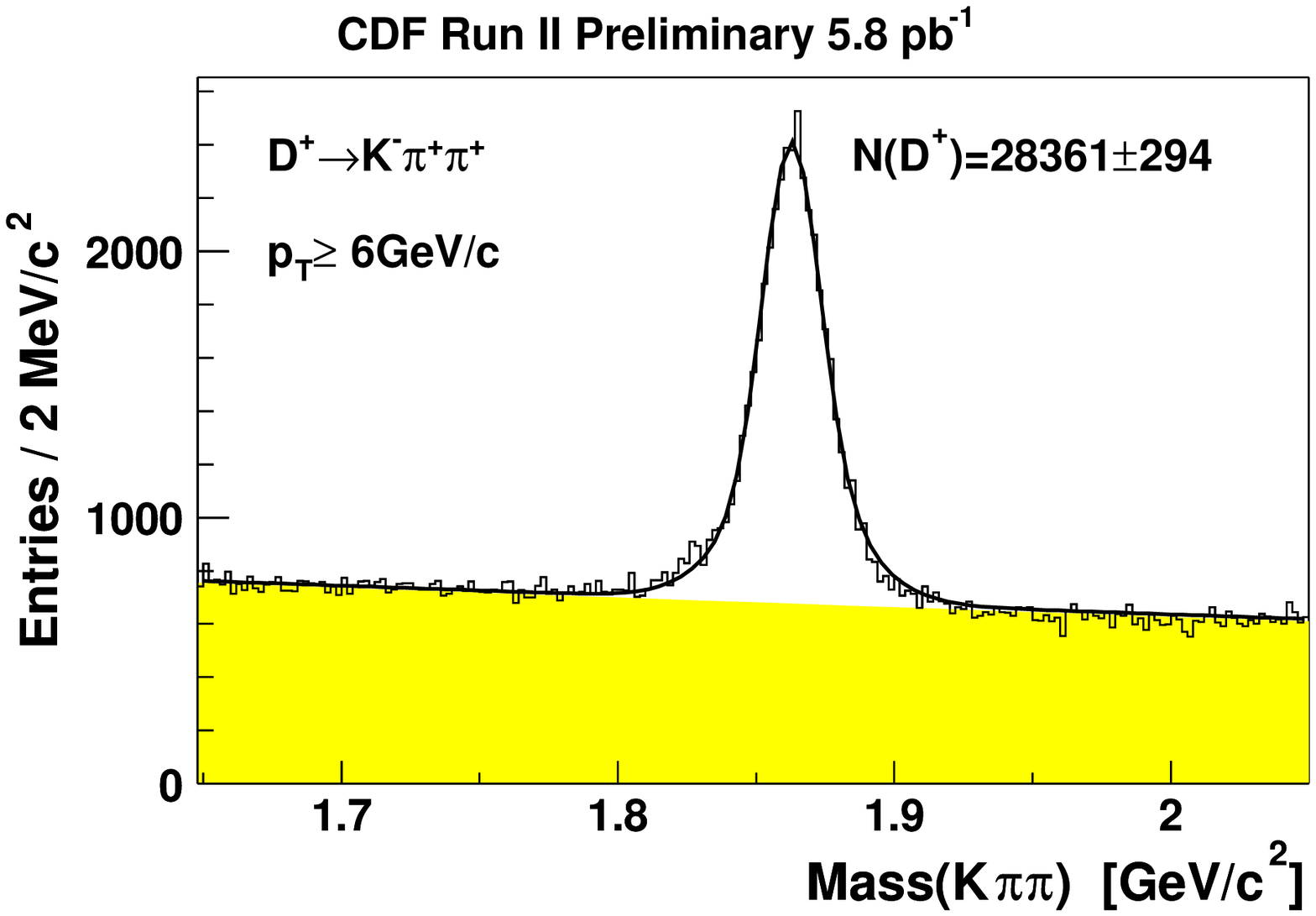}
\epsfxsize=6.1cm
\epsffile{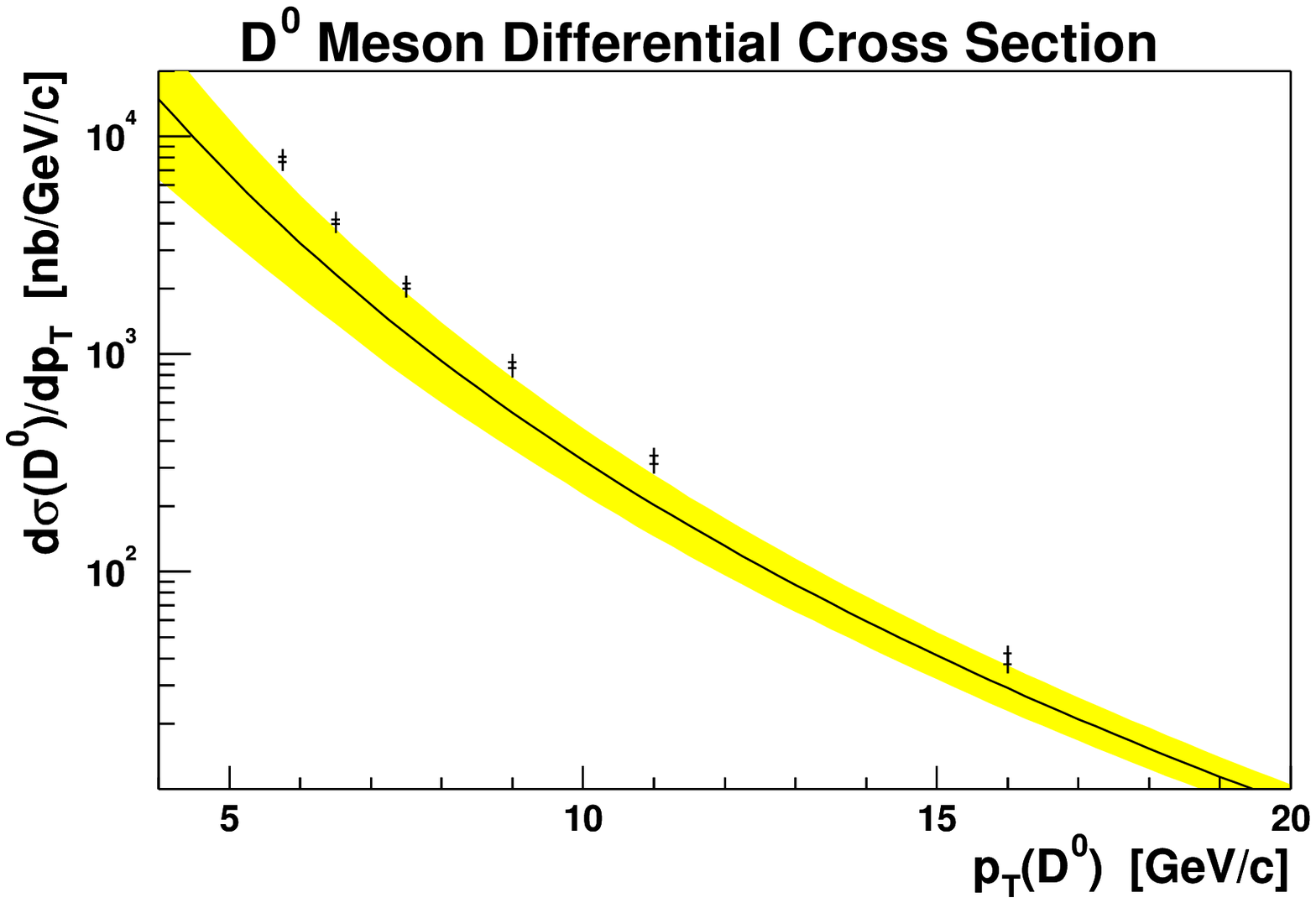}
\put(-205,80){\large\bf (a)}
\put(-30,90){\large\bf (b)}
}
\caption{
(a) Yield for $D^+\rightarrow K^-\pi^+\pi^+$ used in the
charm cross section analysis.
(b) The measured differential cross section for prompt 
$p\bar{p}\rightarrow
D^0 X$, with $|y(D^0)|<1$.  The $D^0$ hadrons arising from $B$ decays
have been removed.  The NLO calculation is from Ref.~\cite{nlo}.
}
\label{fig:charmxs}
\end{figure}

Previously published measurements of the $b$~production
cross section at the Tevatron have consistently been 
higher than the Next-to-Leading-Order (NLO) QCD predictions.  Although
the level of discrepancy
has been reduced with recent theoretical activity, it is not yet clear that
the entire scope of the problem is understood.  Both collider experiments
will again measure the $b$ and $b\bar{b}$ cross 
sections in Run\,II.
To further shed light on this problem, CDF has recently
presented a measurement of the charm production cross
section~\cite{charmxsec}.
Using the secondary vertex trigger, CDF has been able to
reconstruct very large samples of charm decays.  
Figure~\ref{fig:charmxs}(a) shows a fully 
reconstructed $D^+\rightarrow K^-\pi^+\pi^+$ signal comprising almost 30k
events using
5.8~pb$^{-1}$ of data from the beginning of Run\,II.  

Since the events are accepted based upon daughter tracks
with large impact parameter, the sample 
of reconstructed charm decays contains charm from direct $c\bar
c$~production as well as charm from $B$~hadron decays $b\ra c$.
To extract the charm meson cross section,
it is necessary to extract the fraction of $D$ mesons that
are coming from prompt charm production and remove the fraction from
$b\ra c$~decays.  This is done by 
measuring the impact parameter of the charm meson.  If it 
arises from direct $c\bar{c}$ production, the charm 
meson will have a small impact parameter pointing back
to the primary $p\bar p$~interaction vertex. 
If the charm meson 
originates from $B$~decays, it will typically 
not extrapolate back to the primary vertex.
Using this technique, along with a sample of $K^0_S\rightarrow \pi^+ \pi^-$
decays for calibration, CDF finds that 80-90\% of the 
charm mesons originate from direct charm production.  The shorter
charm lifetime is more than compensated by the copious charm
production in hadronic collisions.

The full analysis includes measurements of the differential
cross sections for prompt $D^0$, $D^+$, $D^{*+}$ and $\Ds$ meson
production. The integrated cross section 
results for rapidity $|y|\le 1$ are 
$\sigma(D^0, p_T\ge 5.5~\gevc) = (13.3\pm0.2\pm1.5)$~pb,
$\sigma(D^{*+}, p_T\ge 6.0~\gevc) = (5.2\pm0.1\pm0.8)$~pb,
$\sigma(D^+, p_T\ge 6.0~\gevc) = (4.3\pm0.1\pm0.7)$~pb and
$\sigma(\Ds, p_T\ge 8.0~\gevc) = (0.75\pm0.05\pm0.22)$~pb, respectively.
Figure~\ref{fig:charmxs}(b) shows the comparison between data and
a NLO calculation for the differential 
$D^0$~cross section from Ref.~\cite{nlo}.  The trend seen
in this figure is the same for the other $D$~species.  The 
prediction seems to follow the measured cross section in 
shape, but the absolute cross section is low compared to the
measured results.  This difference in magnitude between the
measured and predicted charm meson cross section is similar to
the difference between data and theory seen in the $B$~meson cross sections.

\subsection{Hadronic Branching Ratios}

\subsubsection{Two-body Charmless $B$ Decays}

With the new SVT trigger, CDF has begun to measure 
$B$~decays with non-leptonic final states.  One set of modes of
particular interest are rare charmless two-body decays as they are potential
modes for $CP$~violation measurements.  Requiring
the final state to consist of two charged hadrons ($B\ra hh$), 
the following modes can be accessed:
$B^0\ra\pi^+\pi^-$, $B^0\ra K^{\pm}\pi^{\mp}$, $\Bs\ra
K^{\pm}\pi^{\mp}$, and $\Bs\ra K^+K^-$. 
The $B^0$ states are also reconstructed at the $e^+e^-$ $B$~factories, 
but the $\Bs$ modes are exclusive to the Tevatron.

\begin{figure}[tbp]
\centerline{
\epsfxsize=6.1cm
\epsffile{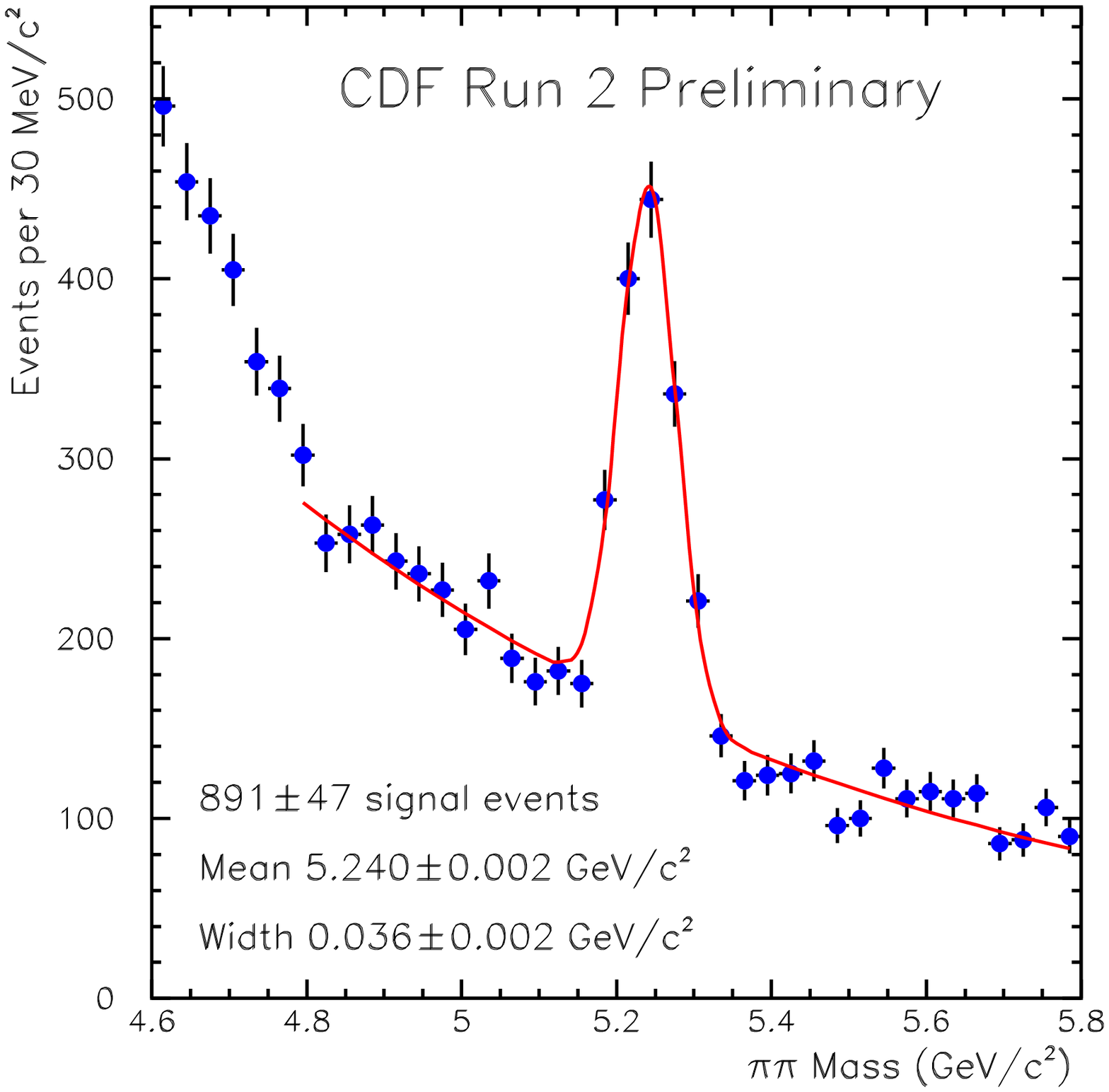}
\epsfxsize=6.1cm
\epsffile{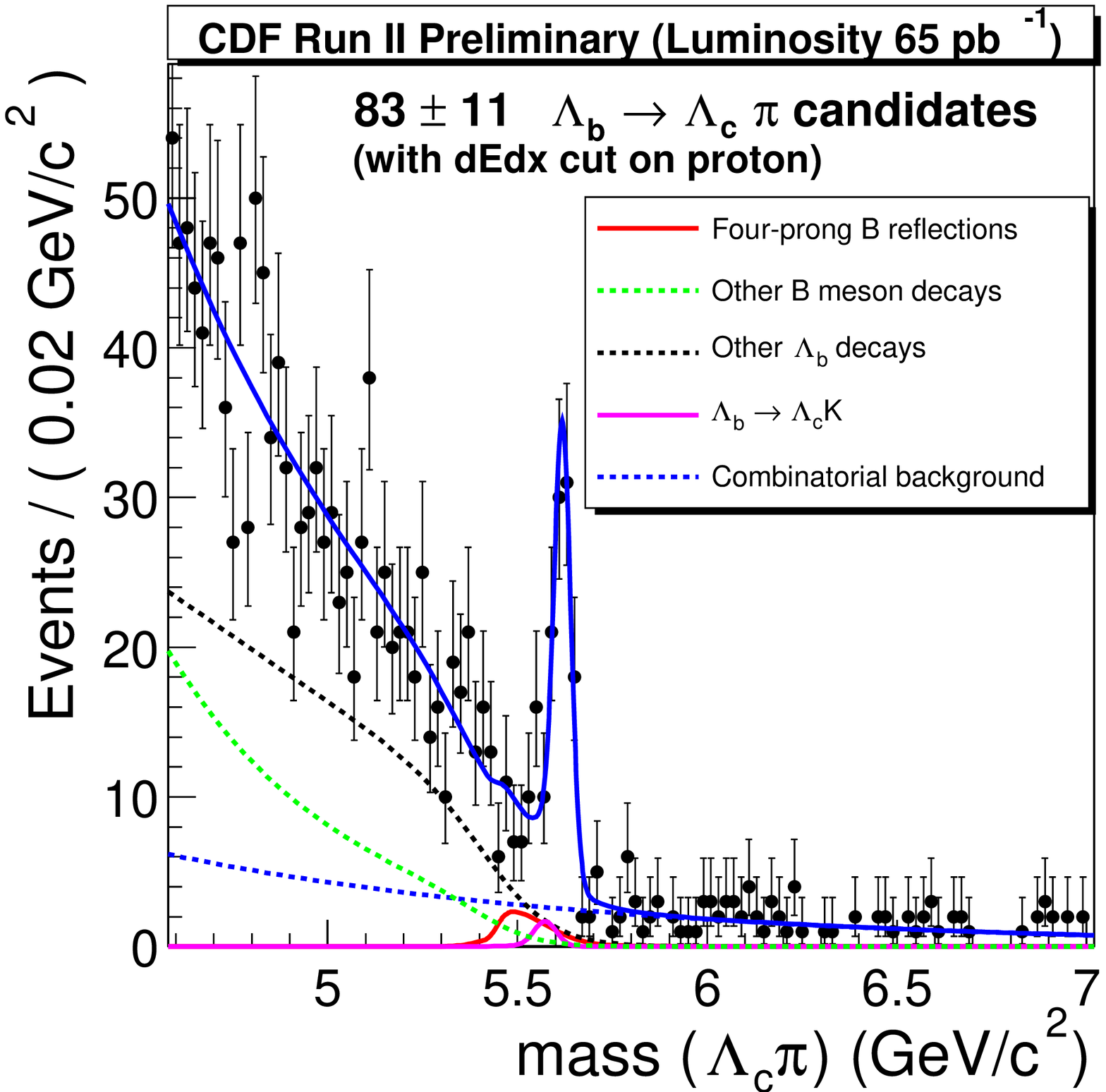}
\put(-210,135){\large\bf (a)}
\put(-30,70){\large\bf (b)}
}
\caption{
(a) The reconstructed two-body $B\rightarrow hh$ 
($h=\pi,K$)
sample from CDF assuming $M_{\pi}$ for both particles.  
(b) The CDF fully reconstructed $\Lambda_b\rightarrow \Lambda_c \pi$
with $\Lambda_c \rightarrow pK^-\pi^+$.
}
\label{fig:bhad}
\end{figure}

Although the branching fractions are small, the final state has two
high-$p_T$ tracks that are relatively efficient for the
displaced-track trigger.
Figure~\ref{fig:bhad}(a) shows the reconstructed signal where all
tracks are assumed to have the pion mass. The
luminosity used for this distribution is 190~pb$^{-1}$.  
A clear signal is 
seen but the width of the peak is significantly larger (41~\mevcc)
than the intrinsic detector resolution.  
This is due to the overlap of the invariant mass distributions from
the four decay modes.
To extract the relative contributions, kinematic information
and d$E$/d$x$ particle identification is used. The particle identification
is calibrated from a large sample of $D^{*+}\rightarrow D^0\pi^+$ decays,
with $D^0\rightarrow K^-\pi^+$.  The charge of the pion from the $D^*$
uniquely identifies the kaon and pion, providing an excellent calibration
sample for the d$E$/d$x$ system.  Although the $K$-$\pi$ separation
is only $\sim\!1.3\,\sigma$, 
this is sufficient to extract the various two-body $B$~decay contributions.

Based on the derived fractions, CDF finds the ratio of branching ratios
$$\frac{{\cal B}(B^0\ra\pi^+\pi^-)}{{\cal B}(B^0\ra K^{\pm}\pi^{\mp})}
= 0.26\pm0.15\pm0.055,$$
and measures the $CP$~asymmetry in the $B^0\ra K\pi$ decay mode to be 
$$
A_{CP} = \frac{N(\bar{B}^0\ra K^-\pi^+) - N(B^0\ra
  K^+\pi^-)}{N(\bar{B}^0\ra K^-\pi^+) + N(B^0\ra K^+\pi^-)}
= 0.02\pm0.15\pm0.017. \nonumber
$$
This analysis includes the first observation of the $\Bs\rightarrow
K^+K^-$ decay mode.  Turning the observed yields into a ratio of branching
ratios, CDF obtains
$$\frac{{\cal B}(\Bs\ra K^+K^-)}{{\cal B}(B^0\ra
  \pi^{+}\pi^{-})} = 2.71\pm0.73\pm0.35\pm0.81,$$
where the systematic uncertainty on $f_s / f_d$ is separately included as the
last error. 

\subsubsection{$\Lambda_b \rightarrow \Lambda_c \pi$ Branching Ratio}

Using the SVT trigger, CDF has also measured
purely hadronic $b$-baryon states. Figure~\ref{fig:bhad}(b) shows a
clean signal of the decay $\Lambda_b \rightarrow \Lambda_c \pi^-$,
with $\Lambda_c \rightarrow pK^-\pi^+$.  The reconstructed
invariant mass plot displays an interesting background structure, with
almost no background above the $\Lambda_b$~peak and a background that
rises steeply going to lower mass. This structure is 
somewhat unique to baryon modes, which are the most massive
weakly decaying $B$~hadron states.  Because the SVT trigger
specifically selects long-lived states, most of the backgrounds
are coming from other heavy flavour ($b$ and $c$) decays.
Since there are no weakly decaying 
$B$~hadrons more massive than the $\Lambda_b$,
there is very little background above the peak.  On the other
hand, going to masses below the peak, lighter $B$~mesons begin to 
contribute.  The background in this mode is growing at lower
masses because there is more phase space for $B^+$, $B^0$,
and $\Bs$~decay modes to contribute.

To extract the number of signal events, $b\bar{b}$
Monte Carlo templates are used to account for the 
reflections seen in the signal window.  The shapes of 
these templates are fixed by the simulation, but their
normalization is allowed to float.
The number of fitted signal events from this distribution is
$96 \pm 13^{+6}_{-7}$.  The primary result
from this analysis is a measurement of the 
$\Lambda_b\rightarrow \Lambda_c \pi^-$ branching ratio relative to
the kinematically similar $B^0\rightarrow D^-\pi^+$ mode.  Taking that 
ratio, along with the PDG~\cite{pdg} values for 
measured branching ratios and production fractions, CDF 
extracts
$
BR(\Lambda_b\rightarrow \Lambda_c \pi^-) =  
(6.5\pm 1.1 \pm 0.9  \pm 2.3)\times 10^{-3},
$
where the errors listed are statistical, systematic and the
final uncertainty originates from the 
errors in the $B^0\rightarrow D^-\pi^+$ branching ratio, the uncertainty on
the ratio of $f_{baryon}/f_d$~fragmentation fractions
and in particular the $\Lambda_c\ra pK\pi$ branching ratio.

\subsection{Rare Decays}

Flavour changing neutral currents (FCNC) are prohibited on tree-level in
the Standard Model (SM).  
Contributions from sources beyond the SM might
measurably enhance the low SM branching fractions of these rare decays.  
This explains the considerable theoretical interest in 
$\Bs \rightarrow \mu^+ \mu^-$ \cite{bsmumutheo}.
In some models, non-SM contributions are large enough to allow an
observation of this decay mode in Run\,II. 

\begin{figure}[tbp]
\centerline{
\epsfxsize=13cm
\epsffile{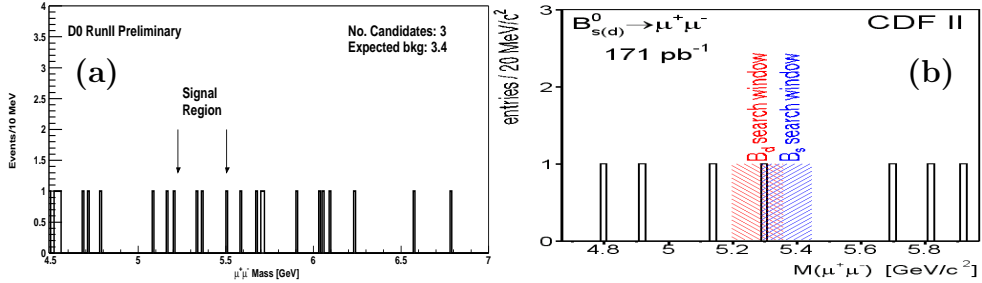}
\put(-345,75){\large\bf (a)}
\put(-30,75){\large\bf (b)}
}
\caption{
Invariant dimuon mass distribution in the search for 
$\Bs \rightarrow \mu^+ \mu^-$ from (a) D\O\ and (b) CDF.
}
\label{fig:bsmumu}
\end{figure}

Experimentally there is considerable background of direct muon pairs in the 
spectrum of reconstructed muons.
This is reduced by requiring the 
two muon tracks to form a displaced vertex and 
selecting candidates with a minimum transverse momentum of 
$p_T>4.0$~\gevc, in addition to the requirement  
that each muon is isolated.
After applying these cuts in the D\O\ analysis, three \Bs~candidates remain
as shown in Figure~\ref{fig:bsmumu}(a). This is consistent
with a background expectation of 3.4 events.
Using the Feldman-Cousins method, D\O\ obtains a
limit on the branching ratio of 
$BR(\Bs \rightarrow \mu^+ \mu^-)<1.6 \cdot 10^{-6}$ at the 90\% confidence
limit. This result is competitive with the CDF Run\,I limit. 

Using data selected with the dimuon trigger, CDF has also searched for the
flavour-changing neutral current decay $\Bs \ra \mu^+\mu^-$.
After applying optimized selection criteria, one event remained in the
$\Bs$ search window as shown in Fig.~\ref{fig:bsmumu}(b). 
This yields an improved upper limit on the branching fraction  of
$9.5\times10^{-7}$ 
($1.2\times10^{-6}$) at the 90\% (95\%) confidence level.
This is more than a factor of two improvement over the previous limit
produced by CDF in Run~I.
In addition,
an upper limit on the branching fraction of $B^0\ra \mu^+\mu^-$ is derived
simultaneously yielding values of $2.5\times10^{-7}$  and
$3.1\times10^{-7}$ at the 90\% and 95\% confidence levels, respectively.

Data selected with the displaced track trigger were used to improve
the limit on the branching fraction of the FCNC decay $D^0 \ra \mu^+\mu^-$.
This search begins by reconstructing a clean sample of the
kinematically similar $D^0 \ra \pi^+\pi^-$ decays using a $D^*$~tag,
followed by muon identification to select $D^0 \ra \mu^+\mu^-$ candidates.
The $D^0 \ra \pi^+\pi^-$ decays serve also as normalization mode.
A new upper limit of $2.5\times10^{-6}$ at the 90\% confidence level
is derived from zero candidates in the search window. It is
almost a factor of two better than the previous best limit.

\section{Conclusion}

After a five year upgrade period, the CDF and D\O\ detectors are back in
operation taking high quality data in Run\,II of the Fermilab Tevatron 
Collider with all subsystems
functional. These include new tracking systems with new central tracking 
devices and silicon vertex detectors for both experiments as well as a
hadronic track trigger for CDF. The understanding of both
detectors is well advanced and first physics results have been presented. 
We also compared different $B$~hadron producers such as
the $\Upsilon(4S)$ with the hadron collider environment and
discussed general features of $B$~physics at a hadron collider.

We reported on the start-up of both Tevatron detectors and presented a
selection of first $B$~physics results from the Tevatron including 
$B$~hadron masses and lifetimes, measurements of the prompt charm cross
section, hadronic branching fractions and rare decays. 
The prospects for one of the ``flagship'' $B$~physics measurements in
Run\,II, the 
observation of \Bs~flavour oscillations, is discussed in detail in a
another contribution to these proceedings~\cite{steffi}. 
More information about various other $B$~physics prospects at the
Tevatron in Run\,II can be found in Ref.~\cite{breport}.

\bigskip

{\small 
I like to thank the organizers of this stimulating meeting for an excellent
conference and my colleagues from the CDF and D\O\ collaboration for their
help in preparing this talk and the proceedings. This work was supported by the
U.S.~Department of Energy under Grant No.~DE-FG02-91ER40682. 
}


\bigskip

\begin{thebibliography}{99}
\bibitem{bfeasi} 
N.~Ellis and A.~Kernan,
Phys.\ Rept.\  {\bf 195} (1990) 23.

\bibitem{cdf_firstB} 
F.~Abe {\it et al.}  [CDF Collaboration],
Phys.\ Rev.\ Lett.\  {\bf 68} (1992) 3403.

\bibitem{myrevart} 
M.~Paulini,
Int.\ J.\ Mod.\ Phys.\ A {\bf 14} (1999) 2791
[hep-ex/9903002].

\bibitem{cdfup} 
R.~Blair {\it et al.}  [CDF-II Collaboration],
{\it ``The CDF-II detector: Technical design report,''}
FERMILAB-PUB-96-390-E (1996).

\bibitem{dup}
S.~Abachi {\it et al.} [The D0 Collaboration],
{\it ``The D0 upgrade: The detector and its physics,''}
FERMILAB-PUB-96-357-E (1996).


\bibitem{charmxsec} 
D.~Acosta {\it et al.}  [CDF Collaboration],
Phys.\ Rev.\ Lett.\  {\bf 91} (2003) 241804 [hep-ex/0307080].

\bibitem{nlo} 
M.~Cacciari and P.~Nason,
JHEP {\bf 0309} (2003) 006
[hep-ph/0306212].

\bibitem{pdg}
K.~Hagiwara {\it et al.}  [Particle Data Group Collaboration], \\
Phys.\ Rev.\ D {\bf 66} (2002) 010001.

\bibitem{bsmumutheo}
A.~Dedes, H.~K.~Dreiner and U.~Nierste,
Phys.\ Rev.\ Lett.\  {\bf 87} (2001) 251804 [hep-ph/0108037]; \\
R.~Arnowitt, B.~Dutta, T.~Kamon and M.~Tanaka,
Phys.\ Lett.\ B {\bf 538} (2002) 121 [hep-ph/0203069]; \\
G.~L.~Kane, C.~Kolda and J.~E.~Lennon
[hep-ph/0310042].

\bibitem{steffi}
S.~Menzemer; these proceedings (2003).

\bibitem{breport}
K.~Anikeev {\it et al.},
FERMILAB-PUB-01-197 (2001) 
[hep-ph/0201071].

\end{thebibliography}
\end{document}